\documentclass[prodmode,acmtoms]{acmsmall}

\acmVolume{0}
\acmNumber{0}
\acmArticle{0}
\acmYear{2014}
\acmMonth{0}

\graphicspath{ {./figures/} }

\usepackage[T1]{fontenc}
\usepackage{amsmath,amssymb,amsfonts,graphicx}
\usepackage{bbm}
\usepackage{epsfig}
\usepackage{epstopdf}
\usepackage{float}
\usepackage[parfill]{parskip}
\usepackage{comment}
\usepackage{cite}
\usepackage{listings}
\usepackage{algorithm}
\usepackage[noend]{algpseudocode}
\usepackage{alltt}
\usepackage{xspace}
\usepackage[usenames,dvipsnames]{xcolor}
\usepackage{hyperref}

\newcommand{\blue}{\textcolor{blue}}
\newcommand{\green}{\textcolor{green}}

\usepackage{tikz}
\usepackage{pgflibraryshapes}
\usetikzlibrary{decorations.pathreplacing}
\usetikzlibrary{backgrounds}
\usetikzlibrary{arrows}
\usetikzlibrary{trees,quadtree}

\def\PETSc{{\sc PETS}c\xspace}
\def\DMPlex{{\tt DMPlex}\xspace}
\def\SF{{\tt SF}\xspace}
\def\Section{{\tt Section}\xspace}
\def\Label{{\tt Label}\xspace}
\def\IS{{\tt IS}\xspace}
\def\Vec{{\tt Vec}\xspace}

\def\function#1{{\tt #1()}\xspace}

\def\cone{\mathbbm{cone}}
\def\supp{\mathbbm{supp}}
\def\cl{\mathbbm{cl}}
\def\st{\mathbbm{st}}
\def\clloc{\mathbbm{cl}_{\mathrm{loc}}}
\def\stloc{\mathbbm{st}_{\mathrm{loc}}}
\def\adj{\mathbbm{adj}}

\begin{document}


\markboth{Knepley \and Lange \and Gorman}{Unstructured Overlapping Mesh Distribution in Parallel}

\title{Unstructured Overlapping Mesh Distribution in Parallel}

\author{MATTHEW G. KNEPLEY
\affil{Rice University}
MICHAEL LANGE
\affil{Imperial College London}
GERARD GORMAN
\affil{Imperial College London}
}

\begin{abstract}
  We present a simple mathematical framework and API for parallel mesh and data distribution, load balancing, and
overlap generation. It relies on viewing the mesh as a Hasse diagram, abstracting away information such as cell shape,
dimension, and coordinates. The high level of abstraction makes our interface both concise and powerful, as the same
algorithm applies to any representable mesh, such as hybrid meshes, meshes embedded in higher dimension, and overlapped
meshes in parallel. We present evidence, both theoretical and experimental, that the algorithms are scalable and
efficient. A working implementation can be found in the latest release of the PETSc libraries.
\end{abstract}




\acmformat{Matthew G. Knepley, Michael Lange, and Gerard J. Gorman, 2014. Unstructured Overlapping Mesh Distribution in Parallel.}

\begin{bottomstuff}
  MGK acknowledges partial support from DOE Contract DE-AC02-06CH11357 and NSF Grant OCI-1147680.
  ML and GJG acknowledge support from EPSRC grant EP/L000407/1 and the embedded CSE programme of the ARCHER UK National Supercomputing Service (http://www.archer.ac.uk).
  All authors acknowledge support from the Intel Parallel Computing Center program through grants to both the University of Chicago and Imperial College London.

  Authors' addresses: 
  M.G. Knepley, Computational and Applied Mathematics, Rice University, Houston, TX; email: knepley@rice.edu; 
  M. Lange, Imperial College London; email: michael.lange@imperial.ac.uk;
  G.J. Gorman, Imperial College London; email: g.gorman@imperial.ac.uk
\end{bottomstuff}

\maketitle

\section{Introduction}

The algorithms and implementation for scalable mesh management, encompassing partitioning, distribution, rebalancing, and
overlap generation, as well as data management over a mesh can be quite complex. It is common to divide meshes into
collections of entities (cell, face, edge, vertex) of different dimensions which can take a wide variety of forms
(triangle, pentagon, tetrahedron, pyramid, \ldots), and have query functions tailored to each specific form~\cite{ITAPS}.
This code structure, however, results in many different cases, little reuse, and greatly increases the complexity and
maintenance burden. On the other hand, codes for adaptive redistribution of meshes based on parallel partitioning such
as the Zoltan library~\cite{Zoltan06}, usually represent the mesh purely as an undirected graph, encoding cells and
vertices and ignoring the topology. For data distribution, interfaces have been specialized to each specific function
space represented on the mesh. In Zoltan, for example, the user is responsible for supplying functions to pack and
unpack data from communication buffers. This process can be automated however, as in
DUNE-FEM~\cite{DednerKlofkornNolteOhlberger10} which attaches data to entities, much like our mesh points described
below.

\begin{figure}[h]\centering
  \raisebox{0.625in}{\begin{tikzpicture}[line join=round,scale=1.0]
  \tikzstyle{normal}=[draw=blue,fill=blue]
  \tikzstyle{parent}=[draw=red,fill=red]
  \tikzstyle{child}=[draw=green!50!black,fill=green!50!black]
  \def\sp{0.1}
  \def\face{+(0,-1) -- +(0,1)}
  \def\point{circle(\sp * 0.707 * 1cm)};
  \tikzstyle{point}=[circle,normal,inner sep=0cm,
  minimum width=\sp * 0.707 * 2.0cm]
  \tikzstyle{pointchild}=[circle,child,inner sep=0cm,
  minimum width=\sp * 0.707 * 2.0cm]

  \begin{scope}
    \draw (-1.2,0.0) node[point]       (A)       {} node [below] {$A$};
    \draw ( 0.2,0.0) node[point]       (B)       {} node [below] {$B$};
    \draw (-2.5,1.5) node[point]       (a)       {} node [left ] {$a$};
    \draw (-1.5,1.5) node[point]       (b)       {} node [left ] {$b$};
    \draw (-0.5,1.5) node[point]       (e)       {} node [left ] {$e$};
    \draw ( 0.5,1.5) node[point]       (c)       {} node [left ] {$c$};
    \draw ( 1.5,1.5) node[point]       (d)       {} node [left ] {$d$};
    \draw (-1.7,3.0) node[point]       (alpha)   {} node [right] {$\alpha$};
    \draw (-0.9,3.0) node[point]       (beta)    {} node [right] {$\beta$};
    \draw (-0.1,3.0) node[point]       (gamma)   {} node [right] {$\gamma$};
    \draw (+0.7,3.0) node[point]       (delta)   {} node [right] {$\delta$};
    \draw [->,thick,draw=gray] (A) -- (a);
    \draw [->,thick,draw=gray] (A) -- (b);
    \draw [->,thick,draw=gray] (A) -- (e);
    \draw [->,thick,draw=gray] (B) -- (c);
    \draw [->,thick,draw=gray] (B) -- (d);
    \draw [->,thick,draw=gray] (B) -- (e);
    \draw [->,thick,draw=gray] (a) -- (alpha);
    \draw [->,thick,draw=gray] (a) -- (beta);
    \draw [->,thick,draw=gray] (b) -- (alpha);
    \draw [->,thick,draw=gray] (b) -- (gamma);
    \draw [->,thick,draw=gray] (c) -- (beta);
    \draw [->,thick,draw=gray] (c) -- (delta);
    \draw [->,thick,draw=gray] (d) -- (delta);
    \draw [->,thick,draw=gray] (d) -- (gamma);
    \draw [->,thick,draw=gray] (e) -- (beta);
    \draw [->,thick,draw=gray] (e) -- (gamma);
  \end{scope}
\end{tikzpicture}}\qquad \begin{tikzpicture}[line join=round,scale=6.0]
  \tikzstyle{boundary}=[draw=blue,fill=blue]
  \tikzstyle{boundaryparent}=[draw=red,fill=red]
  \tikzstyle{boundarychild}=[draw=green!50!black,fill=green!50!black]
  \tikzstyle{leaf}=[draw=blue!20,fill=blue!20]
  \def\sp{0.1}
  \def\face{+(0,-1) -- +(0,1)}
  \def\vertex{circle(\sp * 0.707 * 1cm)};
  \tikzstyle{vertex}=[circle,boundary,inner sep=0cm,
  minimum width=\sp * 0.707 * 2.0cm]
  \tikzstyle{vertexchild}=[circle,boundarychild,inner sep=0cm,
  minimum width=\sp * 0.707 * 2.0cm]

  \begin{scope}
    \draw[morton order,quadtree scale=(1-\sp)] node {}
    child {[leaf] ( 0.1cm,0cm) -- ( 1cm,-0.9cm) -- ( 1cm,0.9cm) --cycle ( 0.62cm,0cm) node {$A$}}
    child {[leaf] (-0.1cm,0cm) -- (-1cm,-0.9cm) -- (-1cm,0.9cm) --cycle (-0.62cm,0cm) node {$B$}};
    \draw[morton order] node {}
    child {(0cm,0cm) node [vertex] {} node [left] {$\alpha$}
      ( 1,-1) node [vertex] {} node [below] {$\beta$}
      ( 1, 1) node [vertex] {} node [above] {$\gamma$}
      [thick,draw=blue] (0.05cm,-0.05cm) -- (0.9cm,-0.9cm) node [midway, below left] {$a$}
      [thick,draw=blue] (0.05cm,0.05cm) -- (0.9cm,0.9cm) node [midway, above left] {$b$}
    }
    child {
      child {
        ( 1, 1) node [vertex] {} node[right] {$\delta$}
        [thick,draw=blue] (-.85cm,-.85cm) -- (.85cm,.85cm) node [midway, below right] {$c$}
      }
      child [missing]
      child {
        [thick,draw=blue] (-.85cm,+.85cm) -- (.85cm,-.85cm) node [midway, above right] {$d$}
      }
      child [missing]
    }
    ;
    \draw[morton order] node {}
    child {
      [thick,draw=blue] (1,-1+\sp) -- (1,1-\sp) node [midway, left] {$e$}
    }
    ;
  \end{scope}
\end{tikzpicture}
  \caption{A simplicial doublet mesh and its DAG (Hasse diagram).}
  \label{fig:doublet}
\end{figure}
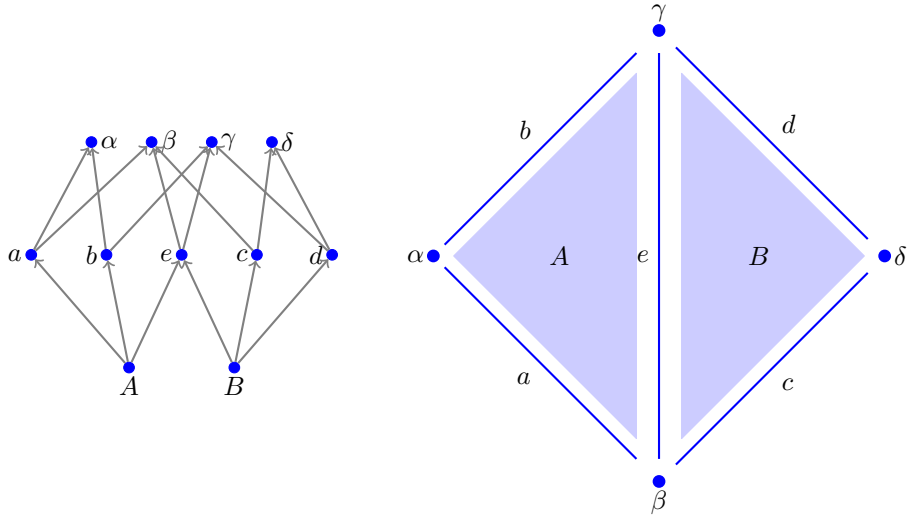

We have previously presented a mesh representation which has a single entity type, called \textit{points}, and a single
antisymmetric relation, called \textit{covering}~\cite{KnepleyKarpeev09}. This structure, more precisely a Hasse
diagram~\cite{Birkhoff1967,HasseDiagram}, can represent any CW-complex~\cite{Hatcher2002,CWcomplex}, and can be
represented algorithmically as a directed acyclic graph (DAG) over the points. It comes with two simple relational
operations, $\cone(p)$, called the \textit{cone} of $p$ or the in-edges of point $p$ in the DAG, and its dual operation
$\supp(p)$, called the \textit{support} of $p$ or the out-edges of point $p$. In addition, we will add the transitive
closure  in the DAG of these two operations, respectively the closure $\cl(p)$ and star $\st(p)$ of point $p$. In Fig.~\ref{fig:doublet}, we show an
example mesh and its corresponding DAG, for which we have $\cone(A) = \{a, b, e\}$ and $\supp(\beta) = \{a, c, e\}$, and
the transitive closures $\cl(A) = \{A, a, b, e, \alpha, \beta, \gamma\}$ and $\st(\beta) = \{\beta, a, c, e, A, B\}$.

In our prior work~\cite{KnepleyKarpeev09}, it was unclear whether simple generic algorithms for \textit{parallel} mesh management tasks could be
formulated, or various types of meshes would require special purpose code despite the generic mesh representation.
Below, we present a complete set of generic algorithms, operating on our generic DAG representation, for parallel mesh
operations, including partitioning, distribution, rebalancing, and overlap generation. The theoretical underpinnings and
algorithms are laid out in Section~\ref{sec:theory}, and experimental results detailed in Section~\ref{sec:results}.

\section{Theory}\label{sec:theory}
\subsection{Overlap Creation}\label{sec:theory_overlap}
We will use the Hasse diagram representation of our computational mesh~\cite{KnepleyKarpeev09}, the \DMPlex class in
\PETSc~\cite{petsc-user-ref,petsc-web-page}, and describe mesh relations (adjacencies) with basic graph operations on a DAG. A
distributed mesh is a collection of closed serial meshes, meaning that they contain the closure of each point,
together with an ``overlap structure'', which marks a subset of the mesh points and indicates processes with which these
points are shared. The default \PETSc representation of the overlap information uses the \SF class, short for
\textit{Star Forest}~\cite{StarForest11}. Each process stores the true owner (root) of its own ghost points (leaves), one side of the
relation above, and construct the other side automatically.

In order to reason about potential parallel mesh algorithms, we will characterize the contents of the overlap using the
mesh operations. These operations will be understood to operate on the entire parallel mesh, identifying shared points,
rather than just the local meshes on each process. To indicate a purely local operation, we will use a subscript,
e.g. $\clloc(p)$ to indicate the closure of a point $p$ evaluated only on the local submesh.

The mesh overlap contains all points of the local mesh adjacent to points of remote meshes in the complete
DAG for the parallel mesh, and we will indicate that point $p$ is in the overlap using an indicator function
$\mathcal{O}$. Moreover, if the overlap contains a point $p$ on a given process, then it will also contain the closure
of $p$,
\begin{equation}
  \mathcal{O}(p) \Longrightarrow \mathcal{O}(q) \qquad \forall q \in \cl(p),
\end{equation}
which shows that if a point is shared, its closure is also shared. This is a consequence of each local mesh being
closed, the transitive closure of its Hasse diagram. We can now examine the effect of increasing the mesh overlap in
parallel by including all the immediately adjacent mesh points to each local mesh.

The set of adjacent mesh point differs depending on the discretization. For example, the finite element method couples
unknowns to all other unknowns whose associated basis functions overlap the support of the given basis function. If
functions are supported on cells whose closure contains the associated mesh point, we have the relation
\begin{equation}
  \adj(p, q) \Longleftrightarrow q \in \cl(\st(p)),
\end{equation}
where we note that this relation is symmetric. For example, a degree of freedom (dof) associated with a vertex is
adjacent to all dofs on the cells containing that vertex. We will call this \textit{FE adjacency}. On the other hand,
for finite volume methods, we typically couple cell unknowns only through faces, so that we have
\begin{equation}
  \adj(p, q) \Longleftrightarrow q \in \supp(\cone(p)),
\end{equation}
which is the common notion of cell-adjacency in meshes, and what we will call \textit{FV adjacency}. This will also be
the adjacency pattern for Discontinuous Galerkin methods.

If we first consider FV adjacency, we see that the cone operation can be satisfied locally since local meshes are
closed. Thus the support from neighboring processes is needed for all points in the overlap. Moreover, in order to
preserve the closure property of local meshes, the closure of that support would also need to be collected.

For FE adjacency, each process begins by collecting the star of its overlap region in the local mesh,
$\stloc(\mathcal{O})$. The union across all processes will produce the star of each point in the overlap region. First,
note that if the star of a point $p$ on the local processes contains a point $q$ on the remote process, then $q$ must be
contained in the star of a point $o$ in the overlap,
\begin{equation}
  q \in \st(p) \Longleftrightarrow \exists o \mid \mathcal{O}(o) \land q \in \st(o).
\end{equation}
There is a path from $p$ to $q$ in the mesh DAG, since $q$ lies in star of $p$, which is the transitive closure. There
must be an edge in this path which connects a point on the local mesh to one 
on the remote mesh, otherwise the path is completely contained in the local mesh. One of the endpoints $o$ of this edge
will be contained in the overlap, since it contains all local points adjacent to remote points in the DAG. In fact, $q$
lies in the star of $o$, since $o$ lies on the path from $p$ to $q$. Thus, the star of $p$ is contained in the union of
the star of the overlap,
\begin{equation}
  \st(p) \in \bigcup_o \st(o).
\end{equation}
Taking the closure of this star is a local operation, since local meshes are closed. Therefore, parallel overlap
creation can be accomplished by the following sequence: each local mesh collects the closure of the star of its overlap,
communicates this to its overlap neighbors, and then each neighbor augments its overlap with the new points. Moreover,
no extra points are communicated, since each communicated point $q$ is adjacent to some $p$ on a remote process.

\subsection{Data Distribution}

We will recognize three basic objects describing a parallel data layout: the \Section~\cite{petsc-user-ref} describing an irregular array of
data and the \SF, StarForest~\cite{StarForest11}, a one-sided description of shared data. A \Section is a map from a domain
of \textit{points} to data sizes, or \textit{ndofs}, and assuming the data is packed it can also calculate an offset for
each point. This is exactly the encoding strategy used in the Compressed Sparse Row matrix
format~\cite{petsc-user-ref}. An \SF stores the owner for any piece of shared data which is not owned by the given
process, so it is a one-sided description of sharing. This admits a very sparse storage scheme, and a scalable algorithm
for assembly of the communication topology~\cite{HoeflerSiebretLumsdaine10}. The third local object, a \Label, is merely
a one-to-many map between integers, that can be manipulated in a very similar fashion to a \Section since the structure
is so similar, but has better complexity for mutation operations.

A \Section may be stored as a simple list of (\textit{ndof}, \textit{offset}) pairs, and the \SF as
(\textit{ldof}, \textit{rdof}, \textit{rank}) triples where \textit{ldof} is the local dof number and \textit{rdof} is
the remote dof number, which means we never need a global numbering of the unknowns. Starting with these two simple
objects, we may mechanically build complex, parallel data distributions from simple algebraic combination
operations. We will illustrate this process with a simple example.

\begin{figure}
\begin{tikzpicture}[scale=1.0]
\path ( 0.0, 0.0) node[anchor=south,text centered] {\bf Point Space};
\path ( 4.0, 0.0) node[anchor=south,text centered] {\bf Dof Space};
\path (8.0, 0.0) node[anchor=south,text centered] {\Section};
\path (12.0, 0.0) node[anchor=south,text centered] {\SF};
\path ( 0.0,  -1.0) node[anchor=south,text centered] {Solution Dofs};
\path ( 4.0,  -1.0) node[anchor=south,text centered] {Adjacent Dofs};
\path (8.0,  -1.0) node (asec) [anchor=south,text centered] {Jacobian Layout};
\path (12.0,  -1.0) node (asf)  [anchor=south,text centered] {Shared Adjacency};
\draw [->] (asec) -- (asf);
\path ( 0.0,  -2.0) node[anchor=south,text centered] {Mesh Points};
\path ( 4.0,  -2.0) node[anchor=south,text centered] {Solution Dofs};
\path (8.0,  -2.0) node (dsec) [anchor=south,text centered] {Data Layout};
\path (12.0,  -2.0) node (dsf)  [anchor=south,text centered] {Shared Dofs};
\draw [->] (dsec) -- (dsf);
\draw [->] (dsf)  -- (asf);
\path ( 0.0,  -3.0) node[anchor=south,text centered] {Mesh Points};
\path ( 4.0,  -3.0) node[anchor=south,text centered] {Mesh Points};
\path (8.0,  -3.0) node (tsec) [anchor=south,text centered] {Topology};
\path (12.0,  -3.0) node (tsf)  [anchor=south,text centered] {Shared Topology};
\draw [->] (tsec) -- (tsf);
\path ( 0.0,  -4.0) node[anchor=south,text centered] {Processes};
\path ( 4.0,  -4.0) node[anchor=south,text centered] {Mesh Points};
\path (8.0,  -4.0) node (psec) [anchor=south,text centered] {Point Partition};
\path (12.0,  -4.0) node (psf)  [anchor=south,text centered] {Shared Points};
\draw [->] (psec) -- (psf);
\draw [->] (psf)  -- (tsf);
\draw [->] (psf)  .. controls (10.0, -3.5) and (10.0, -2.0) .. (dsf);
\path ( 4.0, -5.0) node[anchor=south,text centered] {Processes};
\path (12.0, -5.0) node (nsf) [anchor=south,text centered] {Neighbors};
\draw [->] (nsf) -- (psf);
\end{tikzpicture}
\caption{This figure illustrates the relation between different \Section/\SF pairs. The first column gives the domain
space for the \Section, the second the range space for the \Section and domain and range for the \SF. The \Section and \SF
columns give the semantic content for those structures at each level, and the arrows show how the \SF at each level can
be constructed with input from below. Each horizontal line describes the parallel layout of a certain data set. For
example, the second line down describes the parallel layout of the solution field.}
\label{fig:combine}
\end{figure}
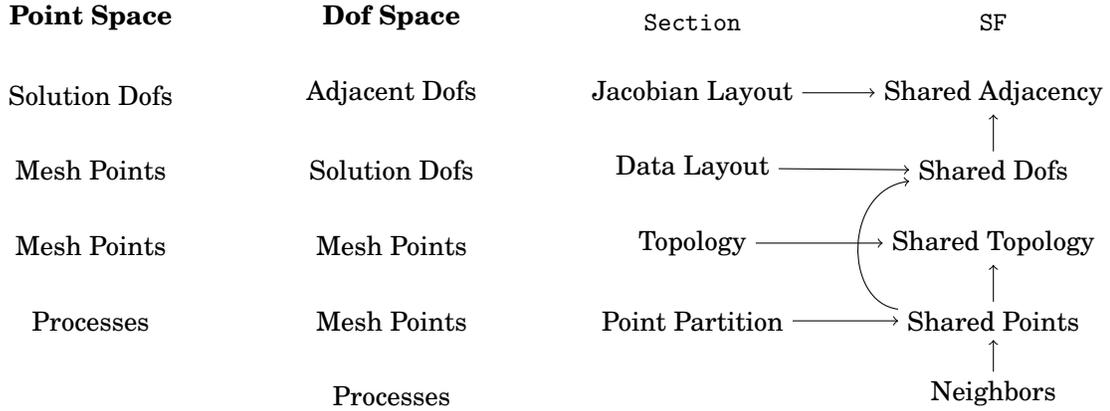

Suppose we begin with a parallel cell-vertex mesh having degrees of freedom on the vertices. On each process, a \Section
holds the number of dofs on each vertex, and an \SF lists the vertices which are owned by other processes. Notice that
the domain (point space) of the \Section is both the domain and the range (dof space) of the \SF. We can combine these two to create a
new \SF whose domain and range (dof space) match the range space of the \Section. This uses the \function{PetscSFCreateSectionSF} function, which is
completely local except for the communication of remote dof offsets, which needs a single sparse broadcast from
dof owners (roots) to dof sharers (leaves), accomplished using \function{PetscSFBcast}. The resulting \SF describes the shared dofs rather than the shared
vertices. We can think of the new \SF as the push-forward along the \Section map. This process can be repeated to
generate a tower of relations, as illustrated in Fig.~\ref{fig:combine}.

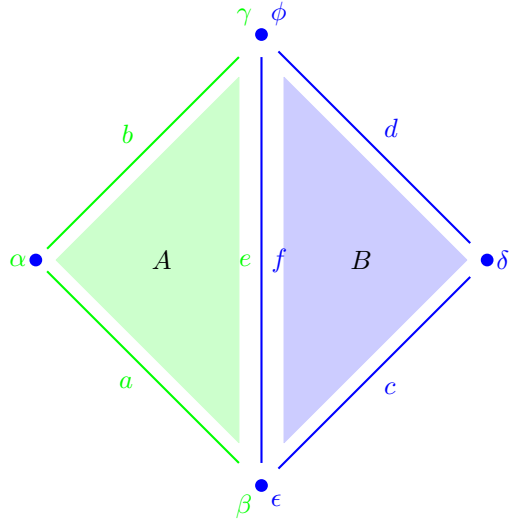
\begin{figure}[h]\centering
  \begin{tikzpicture}[line join=round,scale=6.0]
  \tikzstyle{boundary}=[draw=blue,fill=blue]
  \tikzstyle{boundaryparent}=[draw=red,fill=red]
  \tikzstyle{boundarychild}=[draw=green!50!black,fill=green!50!black]
  \tikzstyle{leaf0}=[draw=green!20,fill=green!20]
  \tikzstyle{leaf1}=[draw=blue!20,fill=blue!20]
  \def\sp{0.1}
  \def\face{+(0,-1) -- +(0,1)}
  \def\vertex{circle(\sp * 0.707 * 1cm)};
  \tikzstyle{vertex}=[circle,boundary,inner sep=0cm,
  minimum width=\sp * 0.707 * 2.0cm]
  \tikzstyle{vertexchild}=[circle,boundarychild,inner sep=0cm,
  minimum width=\sp * 0.707 * 2.0cm]

  \begin{scope}
    \draw[morton order,quadtree scale=(1-\sp)] node {}
    child {[leaf0] ( 0.1cm,0cm) -- ( 1cm,-0.9cm) -- ( 1cm,0.9cm) --cycle ( 0.62cm,0cm) node {$A$}}
    child {[leaf1] (-0.1cm,0cm) -- (-1cm,-0.9cm) -- (-1cm,0.9cm) --cycle (-0.62cm,0cm) node {$B$}};
    \draw[morton order] node {}
    child {(0cm,0cm) node [vertex] {} node [left] {\green{$\alpha$}}
      ( 1,-1) node [vertex] {} node [below left] {\green{$\beta$}} node [below right] {\blue{$\epsilon$}}
      ( 1, 1) node [vertex] {} node [above left] {\green{$\gamma$}} node [above right] {\blue{$\phi$}}
      [thick,draw=blue] (0.05cm,-0.05cm) -- (0.9cm,-0.9cm) node [midway, below left] {\green{$a$}}
      [thick,draw=green] (0.05cm,0.05cm) -- (0.9cm,0.9cm) node [midway, above left] {\green{$b$}}
    }
    child {
      child {
        ( 1, 1) node [vertex] {} node[right] {\blue{$\delta$}}
        [thick,draw=blue] (-.85cm,-.85cm) -- (.85cm,.85cm) node [midway, below right] {\blue{$c$}}
      }
      child [missing]
      child {
        [thick,draw=blue] (-.85cm,+.85cm) -- (.85cm,-.85cm) node [midway, above right] {\blue{$d$}}
      }
      child [missing]
    }
    ;
    \draw[morton order] node {}
    child {
      [thick,draw=blue] (1,-1+\sp) -- (1,1-\sp) node [midway, left] {\green{$e$}} node [midway, right] {\blue{$f$}}
    }
    ;
  \end{scope}
\end{tikzpicture}
  \caption{A parallel simplicial doublet mesh, with points on process 0 blue and process 1 green.}
  \label{fig:doubletParallel}
\end{figure}

We can illustrate the data structures and transformations in Fig.~\ref{fig:combine} by giving concrete examples for the
parallel mesh in Fig.~\ref{fig:doubletParallel}. Given the partition in the figure, we have an \SF $SF_\mathrm{point}$,
called \textit{Shared Points} in Fig.~\ref{fig:combine},
\begin{align}
  SF^0_\mathrm{point} &= \{f \to (e, 1), \epsilon \to (\beta, 1), \phi \to (\gamma, 1)\}, \nonumber\\
  SF^1_\mathrm{point} &= \{e \to (f, 0), \beta \to (\epsilon, 0), \gamma \to (\phi, 0)\}, \nonumber
\end{align}
where the superscript denotes the process on which the object lives. Let us define a data layout for the solution to a
Stokes problem using the Taylor-Hood~\cite{TaylorHood1973} finite element scheme ($P_2$--$P_1$). We define the \Section
$S_u$, called \textit{Data Layout} in Fig.~\ref{fig:combine},
\begin{align}
  S^0_u &= \{c: (2, 0), d: (2, 2), f: (2, 4), \epsilon: (3, 6), \delta: (3, 9), \phi: (3, 12)\} \nonumber\\
  S^1_u &= \{a: (2, 0), b: (2, 2), e: (2, 4), \alpha: (3, 6), \beta: (3, 9), \gamma: (3, 12)\}. \nonumber
\end{align}
Using \function{PetscSFCreateSectionSF}, we obtain a \Section $SF_\mathrm{dof}$, called \textit{Shared Dof} in
Fig.~\ref{fig:combine}, giving us the shared dofs between partitions,
\begin{align}
  SF^0_\mathrm{dof} &= \{4 \to (4, 1), 5 \to (5, 1), 6 \to (9, 1), 7 \to (10, 1), 8 \to (11, 1), \nonumber\\
                   &\quad 12 \to (12, 1), 13 \to (13, 1), 14 \to (14, 1)\} \nonumber
\end{align}
which we note is only half of the relation, and \SF stores one-sided information. The other half which is constructed on
the fly is
\begin{align}
  SF^1_\mathrm{dof} &= \{4 \to (4, 0), 5 \to (5, 0), 9 \to (6, 0), 10 \to (7, 0), 11 \to (8, 0), \nonumber\\
                   &\quad 12 \to (12, 0), 13 \to (13, 0), 14 \to (14, 0)\}. \nonumber
\end{align}

We can use these same relations to transform any parallel data layout into another given an \SF which connects the source
and target point layouts. Suppose that we have an \SF which maps currently owned points to processes which will own them
after redistribution, which we will call a \textit{migration} \SF. With this \SF, we can construct the section after
redistribution and migrate the data itself. This process is show in Alg.~\ref{alg:data_migration}, which
uses \function{PetscSFCreateSectionSF} from above to transform the migration \SF over points to one over dofs, and
also \function{PetscSFDistributeSection} to create the section after redistribution. The section itself can be
distributed using only one sparse broadcast, although we typically use another to setup remote dof offsets
for \function{PetscSFCreateSectionSF}, as shown in Alg.~\ref{alg:section_migration}.

\begin{algorithm}
  \caption{Algorithm for migrating data in parallel}\label{alg:data_migration}
  \begin{algorithmic}[1]
    \Function{MigrateData}{sf, secSource, dtype, dataSource, secTarget, dataTarget}
      \State \Call{PetscSFDistributeSection}{sf, secSource, remoteOff, secTarget}
      \State \Call{PetscSFCreateSectionSF}{sf, secSource, remoteOff, secTarget, sfDof}
      \State \Call{PetscSFBcast}{sfDof, dtype, dataSource, dataTarget}
  \EndFunction
  \end{algorithmic}
\end{algorithm}

\begin{algorithm}
  \caption{Algorithm for migrating a \Section in parallel}\label{alg:section_migration}
  \begin{algorithmic}[1]
    \Function{DistributeSection}{sf, secSource, remoteOff, secTarget}
      \State <Calculate domain (chart) from local SF points>
      \State \Call{PetscSFBcast}{sf, secSource.dof, secTarget.dof} \Comment{Move point dof sizes}
      \State \Call{PetscSFBcast}{sf, secSource.off, remoteOff}     \Comment{Move point dof offsets}
      \State \Call{PetscSectionSetUp}{secTarget}
  \EndFunction
  \end{algorithmic}
\end{algorithm}

These simple building blocks can now be used to migrate all the data for a \DMPlex object, representing an unstructured
mesh of arbitrary dimension composed of cells, each of which may have any shape. The migration of cone
data, coordinates, and labels all follow the general migration algorithm above, since each piece of data can be
expressed as the combination of a \Section, giving the layout, and an array storing the values, in PETSc a \Vec or \IS object. Small
differences from the generic algorithm arise due to the nature of the stored data. For example, the
cone data must also be transformed from original local numbering to the new local numbering, which we accomplish by
first moving to a global numbering and then to the new local numbering using two local-to-global renumberings. After moving the data, we
can compute a new point \SF using Alg.~\ref{alg:sf_migration}, which uses a reduction to compute the unique owners of all
points.

\begin{algorithm}
  \caption{Algorithm for migrating a mesh in parallel}\label{alg:mesh_migration}
  \begin{algorithmic}[1]
    \Function{Migrate}{dmSource, sf, dmTarget}
      \State \Call{ISLocalToGlobalMappingApplyBlock}{l2g, csize, cones, cones}
      \State   \Comment{Convert to global numbering}
      \State \Call{PetscSFBcast}{sf, l2g, l2gMig}                         \Comment{Redistribute renumbering}
      \State \Call{DMPlexDistributeCones}{dmSource, sf, l2gMig, dmTarget}
      \State \Call{DMPlexDistributeCoordinates}{dmSource, sf, dmTarget}
      \State \Call{DMPlexDistributeLabels}{dmSource, sf, dmTarget}
  \EndFunction
  \end{algorithmic}
\end{algorithm}

\begin{algorithm}
  \caption{Algorithm for migrating an SF in parallel}\label{alg:sf_migration}
  \begin{algorithmic}[1]
    \Function{MigrateSF}{sfSource, sfMig, sfTarget}
    \State \Call{PetscSFGetGraph}{sfMig, Nr, Nl, leaves, NULL}
    \For{$p \gets 0,Nl$} \Comment{Make bid to own all points we received}
      \State lowners[p].rank  = rank
      \State lowners[p].index = leaves ? leaves[p] : p
    \EndFor
    \For{$p \gets 0,Nr$} \Comment{Flag so that MAXLOC does not use root value}
      \State rowners[p].rank  = -1
      \State rowners[p].index = -1
    \EndFor
    \State \Call{PetscSFReduce}{sfMigration, lowners, rowners, MAXLOC}
    \State \Call{PetscSFBCast}{sfMigration, rowners, lowners}
    \For{$p \gets 0,Nl, Ng = 0$}
      \If {lowners[p].rank != rank}
        \State ghostPoints[Ng]        = leaves ? leaves[p] : p
        \State remotePoints[Ng].rank  = lowners[p].rank
        \State remotePoints[Ng].index = lowners[p].index
        \State Ng++
      \EndIf
    \EndFor
    \State \Call{PetscSFSetGraph}{sfTarget, Np, Ng, ghostPoints, remotePoints}
  \EndFunction
  \end{algorithmic}
\end{algorithm}

\subsection{Mesh Distribution}



Using the data migration routines above, we can easily accomplish sophisticated mesh manipulation in PETSc. Thus, we can
redistribute a given mesh in parallel, a special case of which is distribution of a serial mesh to a set of
processes. As shown in Alg.~\ref{alg:mesh_distribution}, we first create a partition using a third party mesh partitioner,
and store it as a label, where the target ranks are label values. We take the closure of this partition in the DAG,
invert the partition to get receiver data, allowing us to create a migration \SF and use the data migration algorithms
above. The only piece of data that we need in order to begin, or bootstrap, the partition process is an \SF which
connects sending and receiving processes. Below, we create the complete graph on processes, meaning that any process
could communicate with any other, in order to avoid communication to discover which processes receive from the
partition. Discovery is possible and sometimes desirable, and will be incorporated in a further update.

\begin{algorithm}
  \caption{Algorithm for distributing a mesh in parallel}\label{alg:mesh_distribution}
  \begin{algorithmic}[1]
    \Function{Distribute}{dm, overlap, sf, pdm}
      \State \Call{PetscPartitionerPartition}{part, dm, lblPart}           \Comment{Partition cells}
      \State \Call{DMPlexPartitionLabelClosure}{dm, lblPart}               \Comment{Partition points}
      \For{$p \gets 0,P$}                                                  \Comment{Create process SF}
        \State remoteProc[p].rank  = p
        \State remoteProc[p].index = rank
      \EndFor
      \State \Call{PetscSFSetGraph}{sfProc, P, P, NULL, remoteProc}
      \State \Call{DMPlexPartitionLabelInvert}{dm, lblPart, sfProc, lblMig}
      \State   \Comment{Convert from senders to receivers}
      \State \Call{DMPlexPartitionLabelCreateSF}{dm, lblMig, sfMig}
      \State   \Comment{Create migration SF}
      \State \Call{DMPlexMigrate}{dm, sfMigration, dmParallel}              \Comment{Distribute DM}
      \State \Call{DMPlexDistributeSF}{dm, sfMigration, dmParallel}         \Comment{Create new SF}
  \EndFunction
  \end{algorithmic}
\end{algorithm}

We can illustrate the migration process by showing how Fig.~\ref{fig:doubletParallel} is derived from
Fig.~\ref{fig:doublet}. We begin with the doublet mesh contained entirely on one process. In the partition phase, we
first create a cell partition consisting of a \Section $S_\mathrm{cpart}$ for data layout and an \IS $\mathrm{cpart}$
holding the points in each partition,
\begin{align}
  S_\mathrm{cpart} &= \{0: (1, 0), 1: (1, 1)\}, \nonumber\\
  \mathrm{cpart}  &= \{B, A\}, \nonumber
\end{align}
which is converted to the equivalent \Label, a data structure better optimized for overlap insertion,
\begin{align}
  L_\mathrm{cpart} = \{0 \to \{B\}, 1 \to \{A\}\}, \nonumber
\end{align}
and then we create the transitive closure. We can express this as a \Section $S_\mathrm{part}$, called \textit{Point
Partition} in Fig.~\ref{fig:combine}, and \IS $\mathrm{part}$ with the partition data,
\begin{align}
  S_\mathrm{part} &= \{0: (4, 0), 1: (7, 4)\}, \nonumber\\
  \mathrm{part}  &= \{B, c, d, \delta, A, a, b, e, \alpha, \beta, \gamma\}, \nonumber
\end{align}
or as the equivalent \Label
\begin{align}
  L_\mathrm{part} = \{0 \to \{B, c, d, \delta\}, 1 \to \{A, a, b, e, \alpha, \beta, \gamma\}\}. \nonumber
\end{align}
The bootstrap \SF $SF_\mathrm{proc}$, called \textit{Neighbors} in Fig.~\ref{fig:combine}, encapsulates the data flow
for migration
\begin{align}
  SF_\mathrm{proc} &= \{0 \to (0, 0), 1 \to (1, 1)\}. \nonumber
\end{align}

We have a small problem in that the partition structure specifies the send information, and for an \SF we require the
receiver to specify the data to be received. Thus we need to invert the partition. This is accomplished with a single call to
\function{DMPlexDistributeData} from Alg.~\ref{alg:data_migration}, which is shown in Alg.~\ref{alg:invert_partition}.
This creates a \Section and \IS with the receive information,
\begin{align}
  S^0_\mathrm{invpart} &= \{0: (4, 0)\} \nonumber\\
  \mathrm{invpart}  &= \{B, c, d, \delta\} \nonumber\\
  S^1_\mathrm{invpart} &= \{0: (7, 0)\} \nonumber\\
  \mathrm{invpart}  &= \{A, a, b, e, \alpha, \beta, \gamma\}. \nonumber
\end{align}
and then we convert them back into a \Label $L_\mathrm{invpart}$. This simple implementation for the complex operation
of partition inversion shows the power of our flexible interface for data movement. Since the functions operate on
generic representations of data (e.g. \Section, \SF), the same code is reused for many different mesh types and
mesh/data operations, and only a small codebase needs to be maintained. In fact, the distribution (one-to-many) and
redistribution (many-to-many) operations are identical except for an initial inversion of the point numbering to obtain
globally unique numbers for cones.

\begin{algorithm}
  \caption{Algorithm for inverting a partition}\label{alg:invert_partition}
  \begin{algorithmic}[1]
    \State \Call{MigrateData}{$SF_\mathrm{proc}$, $S_\mathrm{part}$, MPIU\_2INT, $\mathrm{part}$, $S_\mathrm{invpart}$, $\mathrm{invpart}$}
  \end{algorithmic}
\end{algorithm}

After inverting our partition, we combine $L_\mathrm{invpart}$ and $SF_\mathrm{proc}$ using
\function{DMPlexPartitionLabelCreateSF}, the equivalent of \function{PetscSFCreateSectionSF}, to obtain the \SF for
point migration
\begin{align}
  SF_\mathrm{point} &= \{A \to (A, 1), B \to (B, 0), \nonumber\\
  &a \to (a, 1), b \to (b, 1), c \to (c, 0), d \to (d, 0), e \to (e, 1), \nonumber\\
  &\alpha \to (\alpha, 1), \beta \to (\beta, 1), \gamma \to (\gamma, 1), \delta \to (\delta, 1)\}. \nonumber
\end{align}
In the final step, this \SF is then used to migrate all the (\Section, array) pairs in the \DMPlex, such as cones,
coordinates, and labels, using the generic \function{DMPlexMigrate} function.

\subsection{Overlap Generation}

Following the initial distribution of the mesh, which was solely based
on the partitioner output, the set of overlapping local meshes can now
be derived in parallel. This derivation is performed by each process
computing it's local contribution to the set of overlap points on
neighboring processes, starting from an SF that contains the initial
point sharing. It is important to note here that this approach
performs the potentially costly adjacency search in parallel and that
the search space is limited to the set of points initially shared
along the partition boundary.

The algorithm for identifying the set of local point contributions to
neighboring partitions is based on the respective adjacency
definitions given in section~\ref{sec:theory_overlap}. As illustrated
in Alg.~\ref{alg:create_overlap}, the SF containing the initial
point overlap is first used to identify connections between local
points and remote processes. To add a level of adjacent points, the local
points adjacent to each connecting point are added to a partition label
similar to the one used during the initial migration (see
Alg.~\ref{alg:mesh_distribution}), identifying them as now also connected
to the neighboring process. Once the point donations for the 
first level of cell overlap are defined, further levels can be added
through repeatedly finding points adjacent to the current donations.

\begin{algorithm}
  \caption{Algorithm for computing the partition overlap}\label{alg:create_overlap}
  \begin{algorithmic}[1]
    \Function{DMPlexCreateOverlap}{dm, overlap, sf, odm}
    \State \Call{DMPlexDistributeOwnership}{dm, sf, rootSection, rootRank} \Comment{Derive sender information from SF}
    \For{$leaf \gets sf.leaves$}       \Comment{Add local receive connections}
    \State \Call{DMPlexGetAdjacency}{sf, leaf.index, adjacency}
    \For{$a \gets adjacency$}
    \State \Call{DMLabelSetValue}{lblOl, a, leaf.rank}
    \EndFor
    \EndFor
    \For{$p \gets 0,P$}                \Comment{Add local send connections}
    \If{rootSection[p] > 0}
    \State \Call{DMPlexGetAdjacency}{sf, p, adjacency}
    \For{$a \gets adjacency$}
    \State \Call{DMLabelSetValue}{lblOl, a, rootRank[p]}
    \EndFor
    \EndIf
    \EndFor
    \For{$n \gets 1,overlap$}          \Comment{Add further levels of adjacency}
    \State \Call{DMPlexPartitionLabelAdjacency}{lblOl, n}
    \EndFor
    \EndFunction
  \end{algorithmic}
\end{algorithm}

Having established the mapping required to migrate remote overlap
points, we can derive a migration \SF similar to the one used in
Alg.~\ref{alg:mesh_distribution}. As shown in
Alg.~\ref{alg:distribute_overlap}, this allows us to utilize
\function{DMPlexMigrate} to generate the overlapping local sub-meshes,
provided the migration SF also encapsulates the local point
renumbering required to maintain stratification in the DMPlex
DAG, meaning that cells are numbered contiguously, vertices are numbered contiguously, etc. This graph numbering shift can easily be derived from the \SF
that encapsulates the remote point contributions, thus enabling us to
express local and remote components of the overlap migration in a
single \SF.

\begin{algorithm}
  \caption{Algorithm for migrating overlap points}\label{alg:distribute_overlap}
  \begin{algorithmic}[1]
    \Function{DMPlexDistributeOverlap}{dm, overlap, sf, odm}
    \State \Call{DMPlexCreateOverlap}{dm, sf, lblOl}       \Comment{Create overlap label}
    \State \Call{DMPlexPartitionLabelCreateSF}{dm, lblOl, sfOl} \Comment{Derive migration SF}
    \State \Call{DMPlexStratifyMigrationSF}{dm, sfOl, sfMig}    \Comment{Shift point numbering}
    \State \Call{DMPlexMigrate}{dm, sfMig, dmOl}                \Comment{Distribute overlap}
    \State \Call{DMPlexDistributeSF}{dm, sfMig, dmOl}           \Comment{Create new SF}
    \EndFunction
  \end{algorithmic}
\end{algorithm}

\section{Results}\label{sec:results}

The performance of the distribution algorithms detailed in
Alg.~\ref{alg:mesh_distribution} and ~\ref{alg:distribute_overlap} has
been evaluated on the UK National Supercomputer ARCHER, a Cray XE30
with 4920 nodes connected via an Aries
interconnect~\footnote{\url{http://www.archer.ac.uk/}}. Each node
consists of two 2.7 GHz, 12-core Intel E5-2697 v2 (Ivy Bridge)
processors with 64GB of memory. The benchmarks consist of distributing
a three dimensional simplicial mesh of the unit cube across increasing
numbers of MPI processes (strong scaling), while measuring execution
time and the total amount of data communicated per processor. The mesh
is generated in memory using TetGen~\cite{Si2015,tetgen:homepage} and
the partitioner used is
Metis/ParMetis~\cite{KarypisKumar98,parmetis-web-page}.

\begin{figure}[h]\centering
  \includegraphics[width=0.45\textwidth]{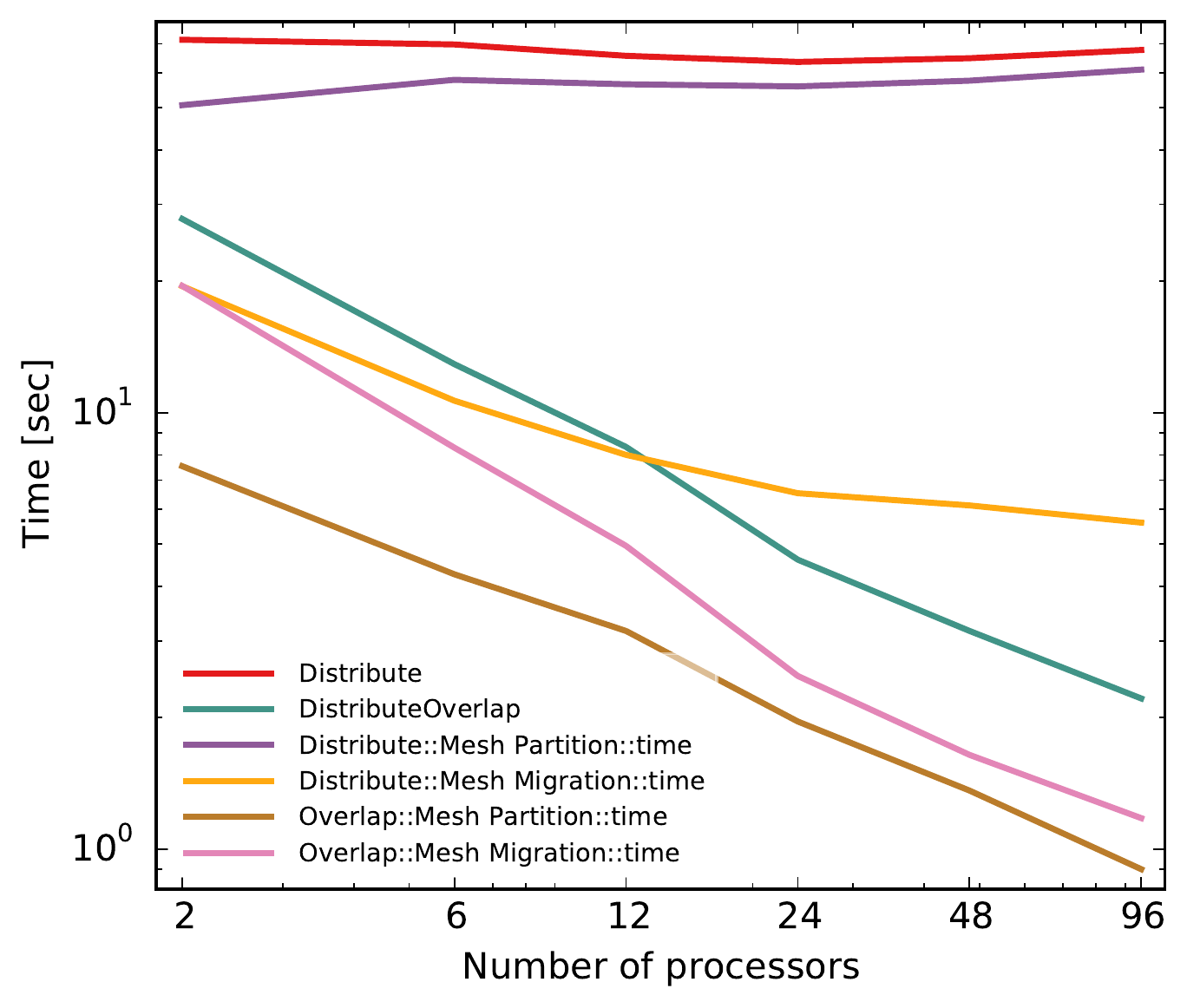}
  \includegraphics[width=0.45\textwidth]{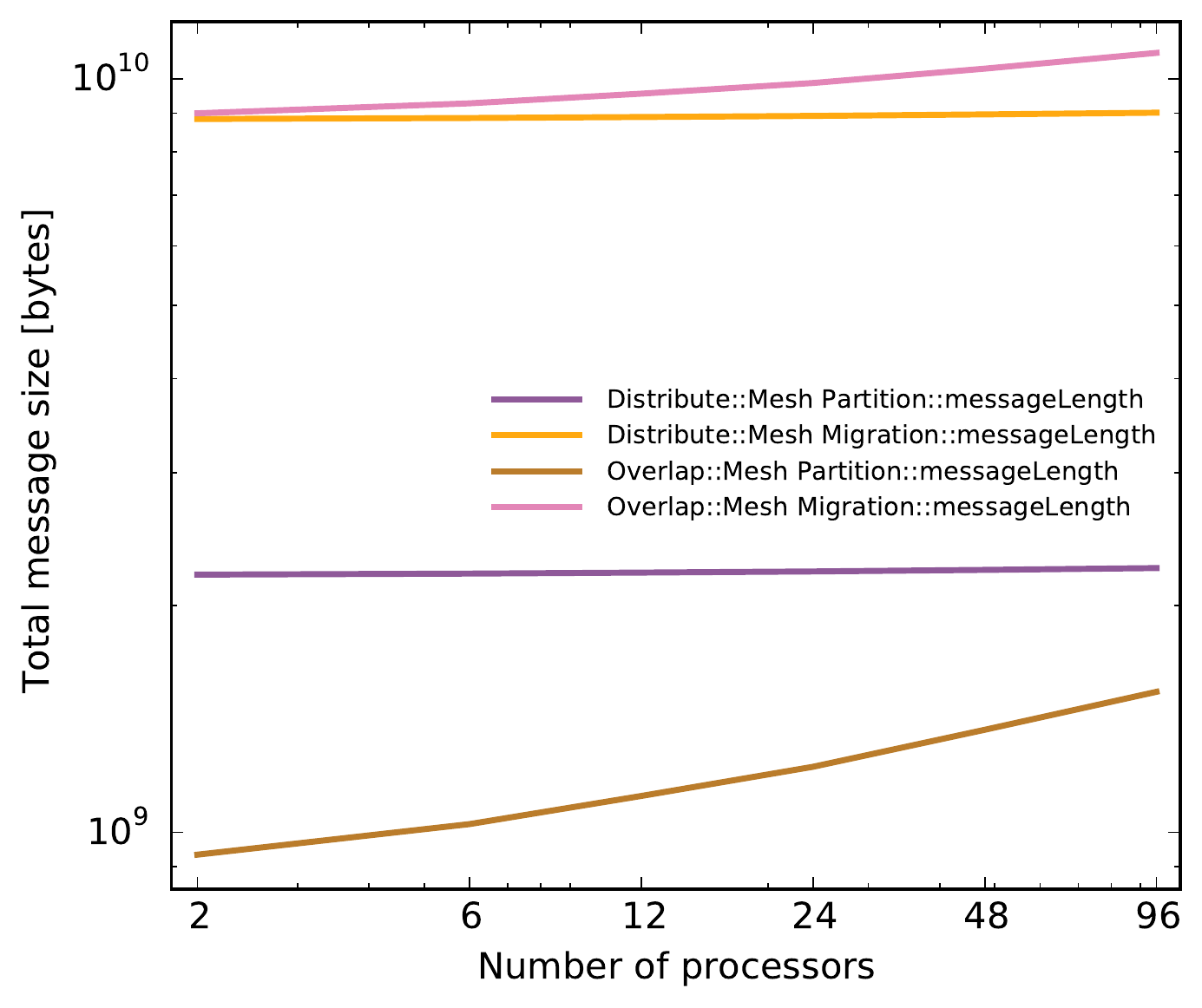}
  \caption{Performance of initial one-to-all mesh distribution of a 3D
    unit cube mesh with approximately 12 million cells. The
    distribution time is dominated by the time to send the serial mesh
    to all processes, and the overlap determination and communication
    time scales linearly with the number of processes.}
  \label{fig:2d_dist}
\end{figure}

The performance of the partitioning and data migration components of
the initial one-to-all mesh distribution, as well as the subsequent
generation of the parallel overlap mapping is detailed in
Fig.~\ref{fig:2d_dist}. The presented run-time measurements indicate
that the parallel overlap generation scales linearly with increasing
numbers of processes, whereas the cost of the initial mesh
distribution increases due to the the sequential partitioning cost.

Data communication was measured as the accumulated message volume
(sent and received) per process for each stage using PETSc's
performance logging~\cite{petsc-user-ref}. As expected, the overall
communication volume during distributed overlap generation increases
with the number of processes due to data replication along the shared
partition boundaries. The communicated data volume during the initial
distribution, however, remains constant, indicating that the increasing
run-time cost is due to sequential processing, not communication of
the partitioning. In fact, the number of high-level communication
calls, such as SF-broadcasts and SF-reductions is constant for meshes
of all sizes and numbers of processes. A model of the total data
volume communicated during the initial distribution of a
three-dimensional mesh can be established as follows:

\begin{align}
  V_{sf} &= 4B * N \nonumber\\
  V_{inversion} &= V_{sf} + 2 * 4B * N \nonumber\\
  V_{stratify} &= V_{sf} + 4B * N \nonumber\\
  V_{partition} &= V_{inversion} + V_{stratify} \nonumber\\
\end{align}

\begin{align}
  V_{cones} &= N_c * 4B * 4 + N_f * 4B * 3 + N_e * 4B * 2 \nonumber\\
  V_{orientations} &= N_c * 4B * 4 + N_f * 4B * 3 + N_e * 4B * 2 \nonumber\\
  V_{section} &= 3 * V_{sf} + 2 * 4B * N \nonumber\\
  V_{topology} &= V_{cones} + V_{orientations} + V_{section} \nonumber\\
  V_{coordinates} &= (3 * 8B + 2 * 4B) * N_v \nonumber\\
  V_{markers} &= 3 * V_{sf} \nonumber\\
  V_{migration} &= V_{topology} + V_{coordinates} + V_{markers}
\end{align}

where $N_c$, $N_f$, $N_e$ and $N_v$ denote the number of cells, faces,
edges and vertices respectively, $N = N_c + N_f + N_e + N_v$ and
$V_{sf}$ is the data volume required to initialize an SF. The unit
square mesh used in the benchmarks has $N_c = 12,582,912$, $N_f =
25,264,128$, $N_e = 14,827,904$, $N_v = 2,146,689$, resulting in
$V_{partition} \approx 1.1GB$ and $V_{migration} \approx 2.8GB$.

%

\begin{figure}[h]\centering
  \includegraphics[width=0.45\textwidth]{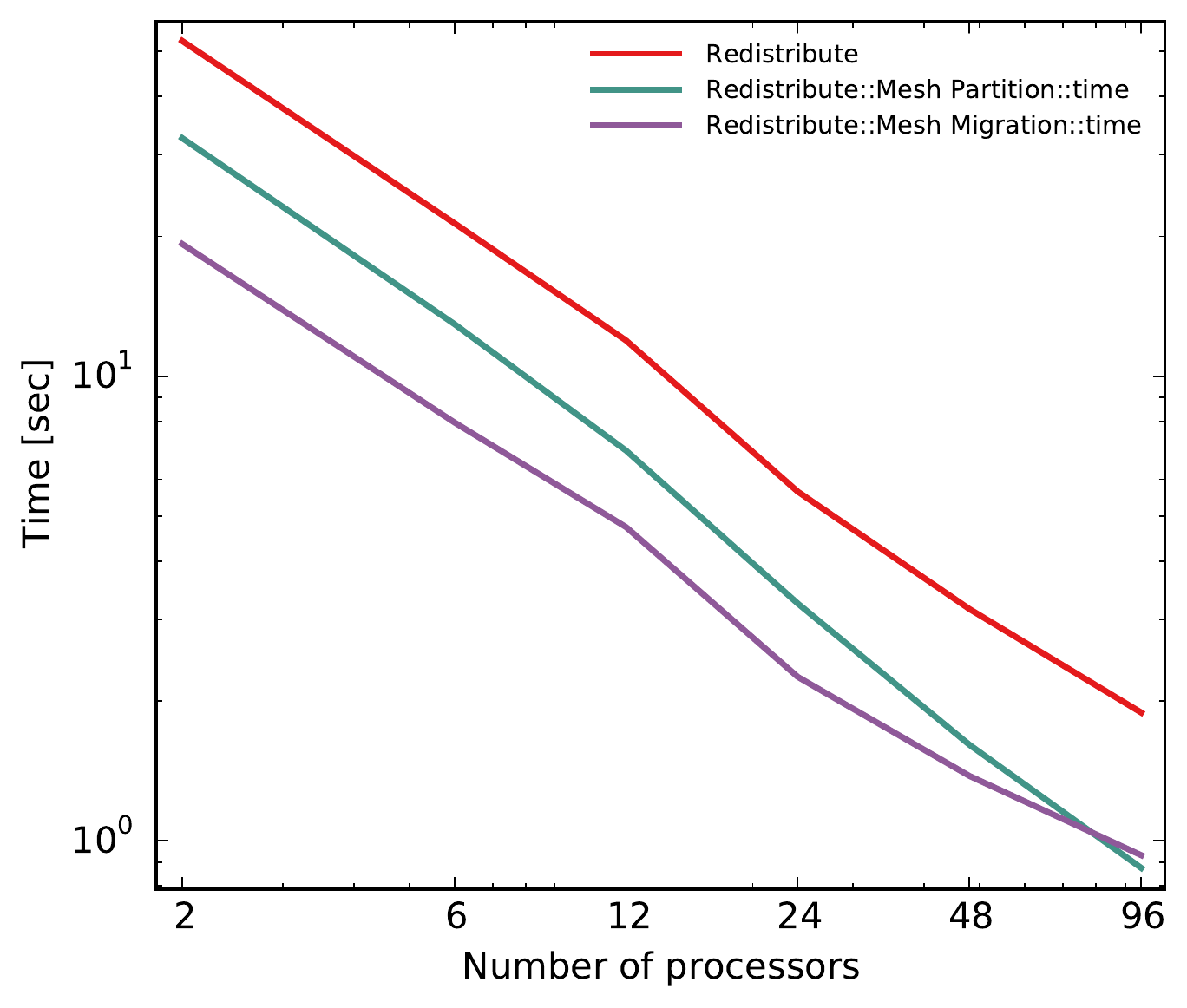}
  \includegraphics[width=0.45\textwidth]{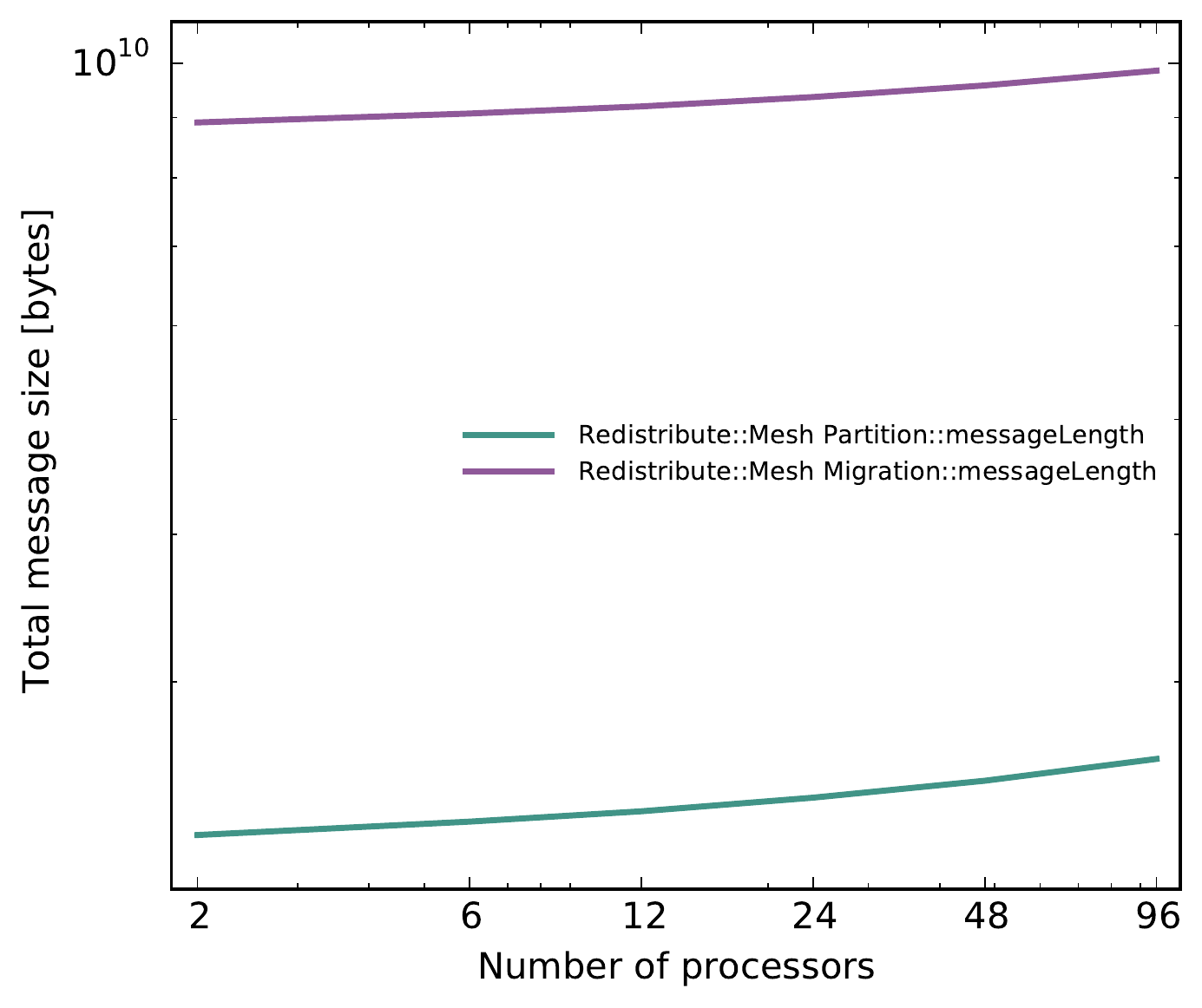}
  \caption{Performance of all-to-all mesh distribution of simplicial
    meshes in 2D and 3D. An initial random partitioning is
    re-partitioned via ParMetis and re-distributed to achieve load
    balancing.}
  \label{fig:2d_redist}
\end{figure}

As well as initial mesh distribution the presented API also allows
all-to-all mesh distribution in order to improve load balance among
partitions. Fig.~\ref{fig:2d_redist} depicts run-time and memory
measurements for such a redistribution process, where an initial bad
partitioning based on random assignment is improved through
re-partitioning with ParMETIS. Similarly to the overlap distribution,
the run-time cost demonstrate good scalability for the partitioning as
well as the migration phase, while the communication volume increases
with the number of processes.

\begin{figure}[h]\centering
  \includegraphics[width=0.7\textwidth]{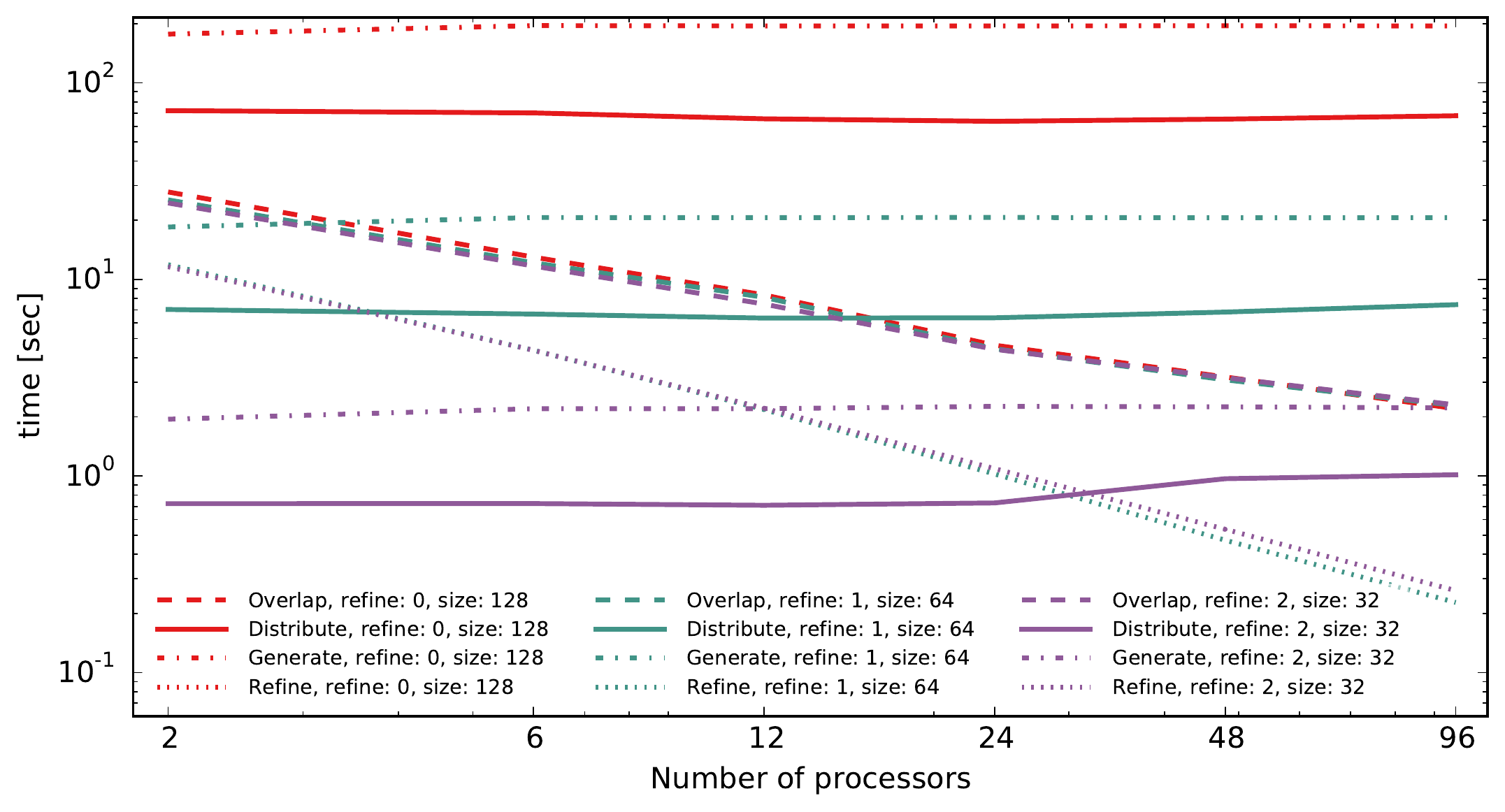}
  \caption{Performance of parallel mesh generation via regular three dimensional refinement.}
  \label{fig:2d_refine}
\end{figure}

As demonstrated in Fig.~\ref{fig:2d_dist}, the sequential overhead of
generating the base mesh on a single process limits overall
scalability of parallel run-time mesh generation. To overcome this
bottleneck, parallel mesh refinement can be used to create
high-resolution meshes in parallel from an initial coarse mesh. The
performance benefits of this approach are highlighted in
Fig.~\ref{fig:2d_refine}, where regular refinement is applied to a
unit cube mesh with varying numbers of edges in each dimension. The
performance measurements show clear improvements for the sequential
components, initial mesh generation and distribution, through the
reduced mesh size, while the parallel refinement operations and
subsequent overlap generation scale linearly. Such an approach is
particularly useful for the generation of mesh hierarchies required
for multigrid preconditioning.

\section{Conclusions}

We have developed a concise, powerful API for general parallel mesh manipulation, based upon the
\function{DMPlexMigrate} capability. With just a few methods, we are able to express mesh distribution from serial to
parallel, parallel redistribution of a mesh, and parallel determination and communication of arbitrary
overlap. Moreover, a user could combine these facilities to specialize mesh distribution for certain parts of a
calculation or for certain fields or solvers, since they are not constrained by a monolithic interface. Moreover, the
same code applies to meshes of any dimension, with any cell shape and connectivity. Thus optimization of these few
routines would apply to the universe of meshes expressible as CW-complexes. In combination with a set of widely used
mesh file format readers this provides a powerful set of tools for efficient mesh management available to a wide
range of applications through PETSc library interfaces~\cite{Lange2015}.

In future work, we will apply these building blocks to the problem of fully parallel mesh construction and
adaptivity. We will input a naive partition of the mesh calculable from common serial mesh formats, and then rebalance
the mesh in parallel. We are developing an interface to the Pragmatic unstructured parallel mesh refinement
package~\cite{rokos2013pragmatic}, which will allow parallel adaptive refinement where we currently use only regular
refinement.

\bibliographystyle{acmsmall}
\bibliography{petsc,petscapp,mesh}

\end{document}